\documentclass[twocolumn,prl,showpacs,preprintnumbers,amsmath,amssymb,superscriptaddress]{revtex4-2}
\usepackage{graphicx}
\usepackage{setspace}

\usepackage{color}
\usepackage{ulem}
\usepackage{enumerate} 
\graphicspath{{./figures/}}

\usepackage{amsmath,amsthm,amssymb,braket}
\usepackage{mathrsfs}

\usepackage[
colorlinks=true, citecolor=blue, urlcolor=blue, linkcolor=blue,
setpagesize=false, bookmarks=false, breaklinks=true
]{hyperref}

\newcommand{\hc}{\hat{c}}

\newcommand{\hH}{\hat{H}}

\newcommand{\hn}{\hat{n}}

\newcommand{\bA}{{\bm A}}
\newcommand{\bE}{{\bm E}}

\newcommand{\eqq}[1]{\begin{align} #1 \end{align}}

\begin{document}

\title{Prominent Dimensional Effects on High-order Harmonic Generation in \\
Strongly Correlated Electron Systems }

\author{Kento Uchida}
 \thanks{K.U., S.T., and Y.M. contributed equally to this work and are co-corresponding authors.}
 \affiliation{Department of Physics, Kyoto University, Sakyo-ku, Kyoto, 606-8502, Japan}
 \affiliation{RIKEN Center for Advanced Photonics, RIKEN 351-0198, Saitama, Japan}
  \email{uchida.kento.4z@kyoto-u.ac.jp}
\author{Shintaro Takayoshi}
 \thanks{K.U., S.T., and Y.M. contributed equally to this work and are co-corresponding authors.}
 \affiliation{Department of Physics, Konan University, Kobe 658-8501, Japan}
  \email{takayoshi@konan-u.ac.jp}
\author{Yuta Murakami}
\thanks{K.U., S.T., and Y.M. contributed equally to this work and are co-corresponding authors.}
\affiliation{Institute for Materials Research, Tohoku University, Sendai, 980-8577, Japan}
\affiliation{RIKEN Center for Emergent Matter Science (CEMS), Wako 351-0198, Japan}
 \email{yuta.murakami@tohoku.ac.jp}
\author{Dongjoon Song}
 \affiliation{Stewart Blusson Quantum Matter Institute, University of British Columbia, Vancouver, BC V6T 1Z4, Canada}
\author{Alannah M. Hallas}
 \affiliation{Stewart Blusson Quantum Matter Institute, University of British Columbia, Vancouver, BC V6T 1Z4, Canada}
\author{Masayuki Watanabe}
 \affiliation{Graduate School of Human and Environmental Studies, Kyoto University, Kyoto 606-8501, Japan}
\author{Takashi Konishi}
 \affiliation{Graduate School of Human and Environmental Studies, Kyoto University, Kyoto 606-8501, Japan}
 \affiliation{Department of Physics, Ritsumeikan University, 
Noji-Higashi 1-1-1, Kusatsu 525-8577, Japan}
\author{Aiko Nakano}
 \affiliation{Department of Physics, Kyoto University, Sakyo-ku, Kyoto, 606-8502, Japan}
\author{Koichiro Tanaka}
 \affiliation{Department of Physics, Kyoto University, Sakyo-ku, Kyoto, 606-8502, Japan}
 \affiliation{RIKEN Center for Advanced Photonics, RIKEN 351-0198, Saitama, Japan}
 \affiliation{HIKARI-COOL Kyoto, Institute for Advanced Study, Kyoto University, Sakyo-ku, Kyoto 606-8501, Japan}

\date{\today}

\begin{abstract}
Dimensionality strongly affects elementary excitations in correlated quantum materials, yet its impact on extreme nonlinear optical responses remains largely unexplored. Here, we combine high-harmonic generation (HHG) experiments on quasi-one-dimensional SrCuO$_2$ and quasi-two-dimensional Pr$_2$CuO$_4$ with nonequilibrium simulations of Hubbard models. We find a pronounced dimensional contrast: SrCuO$_2$ exhibits a robust plateau-like high-harmonic spectrum with weak temperature dependence, whereas Pr$_2$CuO$_4$ shows a monotonic decrease in harmonic yield and strong thermal suppression, especially at higher harmonics. The simulations qualitatively reproduce these trends and identify dimensionality-dependent doublon--holon dephasing, governed by spin--charge coupling, as their microscopic origin. 

These results establish dimensionality and spin--charge coupling as key control parameters for extreme nonlinear optical responses in correlated insulators.
\end{abstract}

\maketitle


\begin{figure*}[t]
 \centering
   \hspace{-0.cm}
   \vspace{0.0cm}
\includegraphics[width=0.9\textwidth]{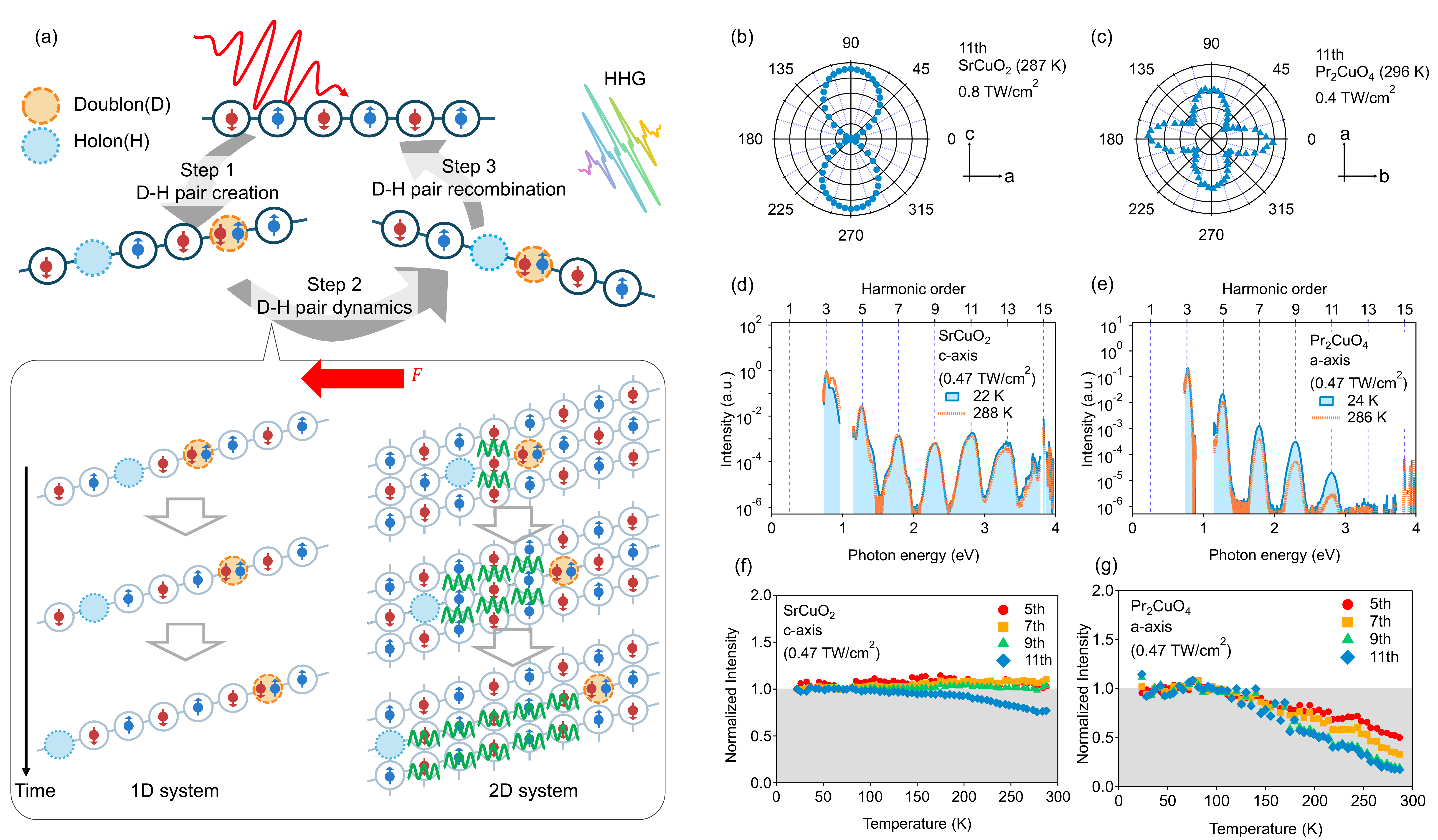} 
 \caption{Left:~(a) Schematic illustration of strong-field charge dynamics in a single-band Mott insulator. The dynamics can be decomposed into three steps: creation, acceleration, and recombination of D-H pairs. The lower panels show the dynamics of D-H pairs driven by an electric field in 1D (left) and 2D (right) systems. In 2D and higher-dimensional systems, the motion of DH pairs can disturb the spin background and generate spin excitations, indicated by green wavy curves. This illustrates the strong influence of spin--charge coupling on the charge dynamics.
 Right:~Experimental results of HHG in $\rm{SrCuO_2}$ and $\rm{Pr_2CuO_4}$. Crystal orientation dependence of 11th harmonic generation in (b) $\rm{SrCuO_2}$ and (c) $\rm{Pr_2CuO_4}$. Typical HH spectra from (d) $\rm{SrCuO_2}$ and (e) $\rm{Pr_2CuO_4}$, where red lines and blue shaded areas correspond to the result at room temperature and around 20 K, respectively. Temperature dependence of normalized HH yields in (f) $\rm{SrCuO_2}$ and (g) $\rm{Pr_2CuO_4}$. The driving field directions in (d)-(g) are along c-axis ($\rm{SrCuO_2}$) and along a-axis ($\rm{Pr_2CuO_4}$), respectively.}
 \label{fig:HHG_Exp}
\end{figure*}
{\it Introduction}--
The interplay between dimensionality and electron correlations gives rise to a wide variety of emergent phases and phenomena in strongly correlated electron systems (SCESs), such as spin liquids and unconventional superconductivity, along with their characteristic elementary excitations~\cite{ImadaRMP1998,Keimer2015Nature}.
Understanding the nature of these elementary excitations and their dependence on dimensionality has been a central issue in characterizing diverse quantum phases. This topic has been extensively investigated under equilibrium or near-equilibrium conditions using a broad range of experimental probes, including transport measurements~\cite{Dressel2005}, linear and nonlinear optical spectroscopy~\cite{Mizuno2000PRB,Kishida2000Nat,Kishida2001PRL,Ashida2002,Moon2008,Boris2011,Basov2011}, and photoemission spectroscopy~\cite{Kim1996PRL,Kim2006NP,Wang2013}.
However, a fundamental question remains far less explored: how do these various elementary excitations respond to electric fields, on ultrafast timescales, in the extreme nonlinear regime?

Addressing this question requires a probe capable of resolving charge dynamics on their intrinsic timescales. In correlated materials, the characteristic energy scales of electron--electron interactions and transfer integrals lie in the electronvolt range, corresponding to femtosecond electronic motion. High-harmonic generation (HHG) in solids, i.e., radiation emitted from charge carriers driven by a strong laser field, provides a natural probe of such ultrafast dynamics~\cite{Ghimire2011NatPhys,Ghimire2019NP,Heide2024NP}. In semiconductors, HHG spectroscopy has been shown to sensitively encode microscopic electronic properties such as band dispersion~\cite{Vampa2015PRL,Luu2015}, Berry curvature and phase~\cite{Luu2018Amorphas,Uzan2024Nat}, and transition-dipole-moment structure~\cite{Uchida2020PRBL}. 

A similar approach can be applied to SCESs, where HHG reflects the ultrafast dynamics of emergent elementary excitations~\cite{Silva2018NatPhoton,Murakami2018PRL,Tancogne-Dejean2018,Ishihara2020,Orthodoxou2021,Murakami2021PRB,Bionta2021PRR,Shao2022PRL,Uchida2022PRL,Alcala2022PNAS,Hansen2022,Kai2023PRB,Nakano2024PRR,Murakami2025PRL,Dziurawiec2026PRB,Ikeda2025arxiV}.
For instance, in single-band Mott insulators, a prototypical class of SCESs, the elementary charge excitations are doublons, i.e., doubly occupied sites, and holons, i.e., empty sites. HHG can then be described by a three-step picture for a doublon--holon (D-H) pair, analogous to that for HHG in gases and semiconductors~\cite{Corkum1993PRL,Lewenstein1994,Vampa2015PRB,Murakami2021PRB}; see Fig.~\ref{fig:HHG_Exp}(a). This process consists of (Step 1) D-H pair creation through tunneling, (Step 2) acceleration of doublons and holons by the electric field, and (Step 3) emission of high-harmonic radiation by the D-H pair recombination [Fig.~\ref{fig:HHG_Exp}(a)]. Importantly, in SCESs, the nature of elementary excitations is sensitive to system parameters such as temperature and dimensionality, reflecting the coupling between different degrees of freedom such as charge, orbital, spin, and lattice. This sensitivity should be encoded in the HHG response, making it a potentially promising probe of elementary excitations in strongly driven regime. 

The main question in this paper is how the effect of spin--charge coupling on the motion of elementary excitations manifests itself in HHG.
A previous HHG experiment on Ca$_2$RuO$_4$, a two-dimensional (2D) Mott insulator, revealed an unconventional enhancement of HHG upon the gap enhancement caused by cooling, a behavior absent in conventional semiconductors~\cite{Uchida2022PRL}. A theoretical analysis of the single-band Hubbard model reproduced this behavior and attributed it to the coupling between charge and spin degrees of freedom, which induces a strong temperature dependence in the dephasing of charge motion~\cite{Murakami2022PRL}.
Importantly, spin--charge coupling is known to depend sensitively on dimensionality. In 1D systems, charge dynamics are expected to be only weakly constrained by the spin background (spin--charge separation), as schematically illustrated in the lower-left panel of Fig.~\ref{fig:HHG_Exp}(a). 
Accordingly, the HHG response is expected to exhibit pronounced differences between 1D and higher-dimensional systems. However, it remains unclear what concrete signatures of this difference appear in HHG, and whether spin--charge separation physics---originally understood as a low-energy phenomenon~\cite{Giamarchi2004book}---can be detected in the strongly driven regime.

To tackle this problem, we combine experimental and theoretical approaches.
Experimentally, we compare HHG from SrCuO$_2$ and Pr$_2$CuO$_4$, and demonstrate that the 1D correlated insulator exhibits stronger nonlinearity and remains robust against thermal fluctuations. 
Theoretically, through the analysis on the single-band Hubbard model, we qualitatively reproduce these features and discuss their relation to distinct strong-field charge dynamics associated with the different roles of spin--charge coupling.
Our results demonstrate that HHG serves as a sensitive probe of elementary charge dynamics in correlated quantum materials.\\

{\it Experiments--}
We experimentally study the cuprates SrCuO$_2$ and Pr$_2$CuO$_4$ as representative quasi-1D and quasi-2D correlated insulators, respectively. SrCuO$_2$ consists of CuO chains along the $c$-axis and has a charge-transfer gap of 1.4 eV~\cite{PopoviPRB2001}. 
The spin degrees of freedom are well described by the $S=1/2$ antiferromagnetic Heisenberg spin-chain with its exchange constant $J\approx\rm{0.2\ eV}$~\cite{MotoyamaPRL1996,ZaliznyakPRL2004}. Angle-resolved photoemission measurements have reported distinct holon–spinon energy splitting as a hallmark of spin–charge separation owing to its quasi-1D nature~\cite{Kim1996PRL,Kim2006NP}. In contrast, $\rm{Pr_2CuO_4}$, while it is also a charge-transfer insulator with a similar optical gap (1.2 eV)~\cite{HomesPRB2002}, exhibits a T'-type planar crystal structure within the $ab$ plane, which reflects its quasi-2D spin and electronic characteristics with an in-plane exchange constant of $J\approx 0.1\ \rm{eV}$~\cite{SumarlinPRB1995,BourgesPRL1997}.

A comparison of the HHG properties in these two systems thus provides an ideal platform for elucidating how dimensionality and spin–charge coupling influence ultrafast charge dynamics in SCESs. We employed mid-infrared (MIR) pulses with a central photon energy of 0.26 eV, far below the optical gap energies, to drive charge carriers in the correlated insulators. To suppress nonlinear propagation effects inside the materials, which can obscure the intrinsic nonlinear optical response, we detected the high-harmonic emission in a reflection geometry. All experimental results presented in this manuscript were obtained at MIR intensities exceeding 0.4 $\rm{TW/cm^2}$ (corresponding to a field strength of 17 MV/cm in vacuum), where non-perturbative nonlinear light-matter interactions dominate, and the physical picture of field-driven charge-packet motion depicted in Fig.~\ref{fig:HHG_Exp}(a) is expected to be applicable.

Figures~\ref{fig:HHG_Exp}(b) and \ref{fig:HHG_Exp}(c) show the crystal orientation dependence of 11th harmonic yields in $\rm{SrCuO_2}$ and $\rm{Pr_2CuO_4}$, respectively. HHG in $\rm{SrCuO_2}$ exhibits a highly anisotropic response, with the maximum HH yield along the $c$-axis and strong suppression along the $a$-axis. This result is consistent with the quasi-1D nature of $\rm{SrCuO_2}$. In contrast, HHG in $\rm{Pr_2CuO_4}$ shows maxima along both the $a$ and $b$ axes, i.e., along the directions connecting nearest-neighbor Cu atoms, which is consistent with its quasi-2D electronic structure. The clear difference in anisotropic response between the two materials shows that HHG is highly sensitive to the geometry of their Cu-O networks.

Figures~\ref{fig:HHG_Exp}(d) and \ref{fig:HHG_Exp}(e) show typical HH spectra from $\rm{SrCuO_2}$ and $\rm{Pr_2CuO_4}$, respectively. In $\rm{SrCuO_2}$, HHG is observed up to the 15th order, which is limited by the detection capability of our experimental setup. The HH spectra exhibit an almost constant yield above the gap energy, forming a remarkable plateau-like structure similar to that observed in HHG from gaseous media. In contrast, the HH spectra of $\rm{Pr_2CuO_4}$ show a monotonic decrease in harmonic yield with increasing harmonic order, and the highest detectable harmonic order is the 11th under the same experimental conditions as those used for $\rm{SrCuO_2}$. Consequently, the HH yield in $\rm{SrCuO_2}$ becomes significantly larger than that in $\rm{Pr_2CuO_4}$ for 9th harmonic generation and above. This contrast between the 1D and 2D systems remains qualitatively unchanged in the strong-field regime beyond the perturbative limit~\cite{SM}.
The much brighter higher-order harmonics observed in 1D correlated insulators than in their 2D counterparts highlight the impact of dimensionality on ultrafast charge dynamics in strongly correlated system under intense laser fields.

The dimensionality-dependent difference between the two materials is also evident in the temperature dependence of HHG, as shown in Figs.~\ref{fig:HHG_Exp}(f) and ~\ref{fig:HHG_Exp}(g).
HHG in $\rm{SrCuO_2}$ is robust against thermal fluctuations and is nearly independent of temperature from the 3rd to the 9th harmonics. In contrast, the HHG in $\rm{Pr_2CuO_4}$ is highly sensitive to thermal fluctuations, exhibiting much stronger suppression with increasing temperature than that observed in $\rm{SrCuO_2}$. In $\rm{Pr_2CuO_4}$, higher harmonics shows stronger suppression, which has same tendency as HHG observed in 2D Mott insulator $\rm{Ca_2RuO_4}$~\cite{Uchida2022PRL}.

{\it Theory--}
To understand the experimentally observed difference in HHG characteristics between the 1D and 2D cuprates, we simulate the nonequilibrium dynamics of the single-band Hubbard model on a 1D chain and a 2D square lattice.
Although the correlated insulating state in cuprates is, strictly speaking, a charge-transfer insulator, its physics is often captured by an effective single-band Hubbard model obtained by downfolding the Cu-$3d$ and O-$2p$ degrees of freedom~\cite{Kishida2001PRL,Shinjo2021PRB,Terashige2019SciAdv}.
To elucidate temperature effects and the role of spin--charge separation, we employ a finite-temperature extension of the infinite time-evolving block decimation (iTEBD) method~\cite{Vidal2003PRL,Vidal2007PRL,Verstraete2004PRL} for the 1D system and time-dependent dynamical mean-field theory (td-DMFT)~\cite{Georges1996,Aoki2013,Eckstein2010,Murakami2025RMP} for the 2D system.
The iTEBD method is based on matrix product states and is well suited for 1D systems , and its accuracy can be systematically controlled by the bond dimension. By contrast, td-DMFT is well suited for describing Mott physics in higher-dimensional systems.
In practice, error accumulation with the time passage prevents us from the long-time simulation in the finite-temperature iTEBD method. Since the time interval is not enough to treat a finite-duration laser pulse, we utilize a continuous-wave (CW) electric field switched on at $t=0$, $E(t)=E_0\sin(\Omega t)$.
The HH spectrum $I_{\rm HHG}(\omega)$ is then evaluated from the current $J(t)$ using a window function spanning a few half-cycles of the field~\cite{SM}.
Using td-DMFT and zero-temperature iTEBD, both of which allow longer-time simulations, we confirm that this protocol gives results qualitatively very similar to those obtained with a conventional finite-duration pulse~\cite{SM}.

\begin{figure}[t]
 \centering
   \hspace{-0.cm}
   \vspace{0.0cm}
\includegraphics[width=87mm]{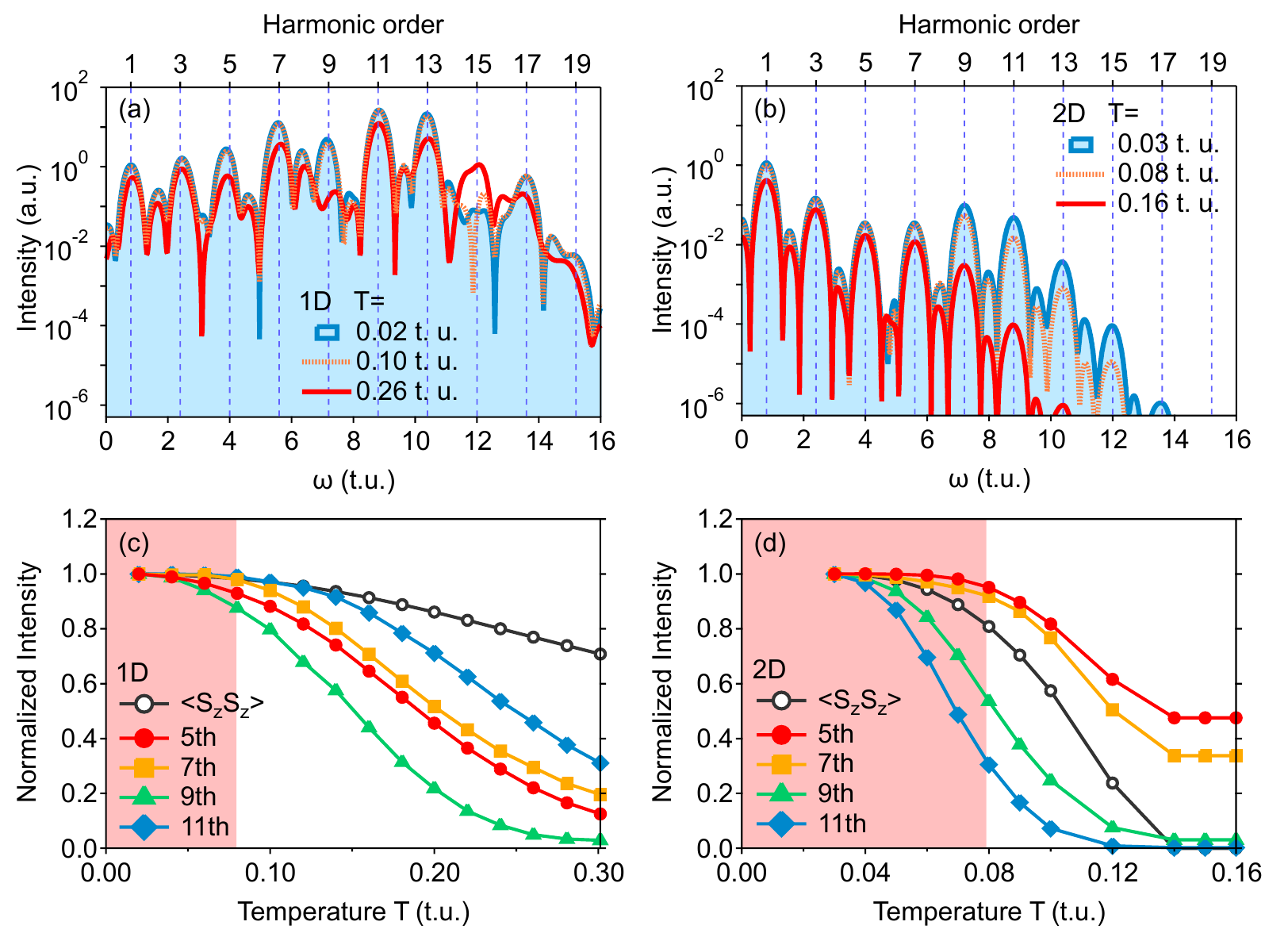} 
 \caption{Comparison of 1D and 2D simulations of HH spectra under CW excitation.
(a), (b) HH spectra $I_{\rm HHG}(\omega)$ for (a) the 1D and (b) the 2D systems.
(c), (d) Temperature dependence of the $n$th-harmonic peak intensity $I_n$ and the nearest-neighbor spin--spin correlation $\langle S_z S_z \rangle$.
The shaded region denotes temperatures up to the room temperature. The HH spectra are evaluated with the box window described in the SM.}
 \label{fig:HHG_Theory}
\end{figure}

To make a reasonable comparison between the 1D and 2D systems within the numerical limitations of iTEBD and td-DMFT, we use common parameters in both cases: the bandwidth $W=4$, the Coulomb interaction $U=7$, and the driving frequency $\Omega=0.8$ in theoretical units.
In these units, we set $\hbar$, the bond length $a$, the electron charge $q$, and the Boltzmann constant $k_{\rm B}$ to unity.
This choice is motivated by the fact that SrCuO$_2$ and Pr$_2$CuO$_4$ have similar gap energies, and we choose the ratio between the Mott gap and the driving frequency to be comparable to that in the experiment. The ratio $U/W$ also lies in the typical range for cuprates. In the following, we fix the driving-field amplitude at $E_0=1.2$, and apply the field along the bond direction.
For rough comparison with the experiment, identifying $\Omega=0.8$ with $0.26$ eV gives an energy unit of $0.325$ eV, corresponding to temperature and time units of approximately $3.8\times10^3$ K and $2.0$ fs. For $a=3.9$ \AA, a typical Cu--Cu distance in cuprates, $E_0=1.2$ corresponds to approximately $10$ MV/cm.

\begin{figure}[t]
 \centering
   \hspace{-0.cm}
   \vspace{0.0cm}
\includegraphics[width=90mm]{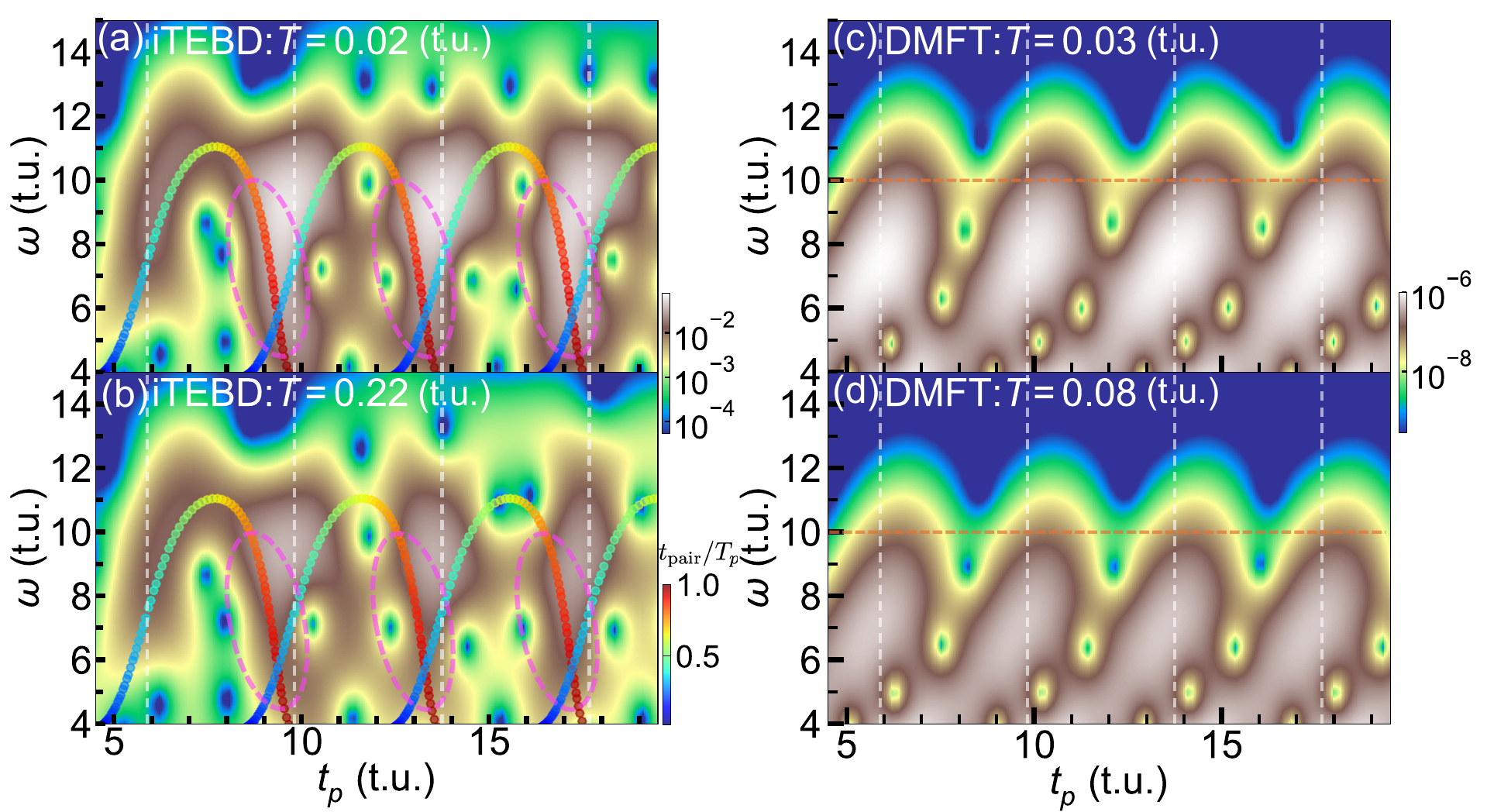} 
 \caption{Comparison between the subcycle spectra of the 1D and 2D simulations under CW excitation.The spectra are shown over two driving cycles, while the characteristic emission pattern repeats every half cycle.
(a,b) Results for the 1D system at (a) $T=0.02$ and (b) $T=0.22$.
The colored dots indicate the emission energy at time $t_p$ predicted by the doublon--holon three-step model.
(c,d) Results for the 2D system at (c) $T=0.03$, (d) $T=0.08$.
For the subcycle analysis, we use a Gaussian window with width $\sigma_p=0.7$.
In panels (b) and (d), the temperatures are chosen to give comparable reductions in the spin--spin correlation $\braket{S_zS_z}$ from the lowest-temperature values in the 1D and 2D systems.
The dashed lines in the different panels are guides to the eye and are placed at the same positions. The white vertical lines indicate the times when $|E(t)|=E_0$.}
 \label{fig:Subcycle_Theory}
\end{figure}

We show the results of calculations in Fig.~\ref{fig:HHG_Theory}.
In the 1D case [Fig.~\ref{fig:HHG_Theory}(a)], the HH spectrum remains nearly flat up to the 13th harmonic, indicating efficient generation of higher-order harmonics. In contrast, in the 2D case [Fig.~\ref{fig:HHG_Theory}(b)], the harmonic yield decreases roughly monotonically with increasing harmonic order.
This difference between the 1D and 2D systems is qualitatively consistent with the experiment, where a clear plateau structure, indicative of strong nonlinearity, is observed in the 1D material, while no plateau is detected in the 2D material.
Strictly speaking, theoretical simulations overestimate the degree of nonlinearity compared with experiments. This discrepancy should be attributed to the absence of additional dephasing and relaxation mechanisms, such as electron--phonon coupling~\cite{Yabana2022PRB,Alvaro2026PRB}, disorder~\cite{Orlando2018}, and the spatially inhomogeneous intensity profile of the driving field~\cite{Floss2018}.

The corresponding subcycle-resolved signals underpin the pronounced difference in the HHG spectra; see Figs.~\ref{fig:Subcycle_Theory}(a) and \ref{fig:Subcycle_Theory}(b) for the results at the lowest temperature.
Here, we evaluate the subcycle spectrum $I(\omega,t_p)$ by applying a Gaussian window centered at time $t_p$ to the current, which provides the time--frequency profile of the radiation within the pump cycle~\cite{SM}.
For the 1D results, we also show the prediction from the D-H three-step model, which incorporates the exact D-H dispersion obtained from the Bethe ansatz~\cite{Murakami2021PRB}.
In 1D, the radiation persists into the later-time region of a half cycle, reflecting the long-lived coherence of D-H pairs, whereas in 2D the radiation is concentrated in the earlier-time region of a half cycle.
These results indicate substantial suppression of D--H coherence in the 2D system, where the motion of a doublon or holon inevitably disturbs the spin background~\cite{Murakami2022PRL,Murakami2025PRL}, causing D-–H dephasing [lower right of Fig.~\ref{fig:HHG_Exp}(a)]. In the 1D system, by contrast, since spin--charge coupling is practically absent even for high-energy excitations at low temperatures, a D–-H pair created via quantum tunneling propagates coherently for a long time, which leads to efficient radiation of higher-order harmonics [lower left of Fig.~\ref{fig:HHG_Exp}(a)].

We now discuss the temperature dependence of the HHG spectra; see Figs.~\ref{fig:HHG_Theory}(c) and (d).
Within the range up to room temperature (the pink shaded area), the simulations reproduce the experimental trends: the temperature dependence is weak in 1D, whereas in 2D the HH intensity is strongly suppressed with increasing temperature, especially for higher-order harmonics.
The latter behavior, as pointed out previously \cite{Murakami2022PRL}, can be attributed to the rapid reduction of D--H coherence by strong spin--charge coupling~\cite{Murakami2022PRL}.
This directly affects Step 2 of the three-step picture, namely D--H pair propagation, and thus more efficiently suppresses higher-order harmonics associated with longer trajectories.
Consistently, the subcycle spectra show that the signal duration becomes shorter as the coherence is reduced; compare Figs.~\ref{fig:Subcycle_Theory}(c) and (d).
In 1D, by contrast, the subcycle spectra change only weakly within this temperature range.

Furthermore, by systematically exploring temperatures beyond the experimental range, we identified a clearer contrast in the temperature dependence of HHG between the 1D and 2D systems.
We find that the HH intensity is also suppressed in the 1D system at higher temperatures despite the expected weak spin-charge coupling.
However, the temperature dependence is qualitatively different from the 2D system.
In the 1D case, the overall shape of the HH spectrum remains nearly unchanged with temperature; see Fig.~\ref{fig:HHG_Theory}(a).
Indeed, the peak intensities from the 3rd to the 11th harmonics exhibit similar temperature dependence, as shown in Fig.~\ref{fig:HHG_Theory}(c).
This is in stark contrast to the 2D system, where the suppression becomes systematically stronger for higher-order harmonics.
In Figs.~\ref{fig:HHG_Theory}(a)(c), we also show the nearest-neighbor spin correlation $\langle S_zS_z\rangle$, which provides a measure for the disturbance of the spin background in the Mott insulating state.
The reduction of $\langle S_zS_z\rangle$ with increasing temperature is much weaker in the 1D system than in the 2D system; nevertheless, a comparable suppression of the HH yield is observed when the reduction of $\langle S_zS_z\rangle$ becomes comparable. This suggests that the spin background can still affect HHG in the high-temperature regime, even in the 1D system. 

Let us now consider how the spin background affects charge motion in 1D. 
Ideally, spin-charge separation allows charges to move without creating spin excitations, as shown in Fig.~\ref{fig:HHG_Exp}(a). Nevertheless, a finite $U$ gives a nonzero spin-exchange coupling, $J_{\rm ex}=4t_{\rm hop}^2/U$, the D--H dispersion therefore still depends on the underlying spin configuration. Since HHG reflects a thermal ensemble of radiation from accelerated D--H pairs in different spin backgrounds for finite temperatures, this dispersion fluctuation causes effective D--H decoherence, explaining the reduced HH yields at elevated temperatures.

However, the overall spin-background-induced dephasing should be much weaker in the 1D system than in the 2D system, because in 2D the charge motion inevitably disturbs the spin background in addition to the mechanism discussed above. 
This raises the question of why comparable reductions in spin correlations lead to comparable HHG suppression but in qualitatively different ways in 1D and 2D.

Comparing the subcycle spectra at different temperatures and in different dimensions clarifies these points; see Fig.~\ref{fig:Subcycle_Theory}.
In 1D at low temperatures, as mentioned above, HHG is dominated by long D-H trajectories with large tunneling--recombination intervals, owing to the long D-H coherence time.
At elevated temperatures, D--H decoherence suppresses the contribution from these long trajectories, but they remain the dominant contribution; see Fig.~\ref{fig:Subcycle_Theory}(b).
Higher harmonics are therefore not necessarily more strongly suppressed, because they are associated with shorter trajectories.
In contrast, in 2D, the dominant contribution comes from short trajectories, and the reduction of coherence more strongly suppresses the relatively longer trajectories responsible for higher harmonics; see Figs.~\ref{fig:Subcycle_Theory}(c) and (d).
We thus attribute the different temperature dependence in 1D and 2D to the dominant D--H trajectories and their sensitivity to spin-background-induced dephasing.
Furthermore, the long D--H trajectories dominant in 1D can be efficiently suppressed even by weak dephasing. This may explain why comparable changes in spin correlations lead to comparable reductions in the HH yield in 1D and 2D, despite the weaker spin-background perturbation in 1D.

{\it Conclusion--} We have clarified the role of dimensionality in HHG from SCESs by combining experiments on 1D and 2D cuprates with numerical simulations of the corresponding Hubbard models. We find that the 1D cuprate exhibits a plateau-like HH spectrum, whereas the HH yield in the 2D cuprate decreases monotonically with harmonic order. This contrast, as well as the distinct temperature dependence of the HH yield, is qualitatively reproduced by the simulations. We identify spin--charge coupling as a key many-body factor governing extreme nonlinear optical responses in Mott insulators.

Importantly, our work demonstrates that 1D Mott systems can sustain enhanced nonperturbative optical nonlinearity, extending the previously known large perturbative nonlinear response~\cite{Kishida2000Nat,Kishida2001PRL} to the high-order regime. Furthermore, the present results suggest that HHG, especially when combined with attosecond metrology, can provide a time-domain probe of many-body charge dynamics and spin--charge coupling in SCESs.

\begin{acknowledgments}
{\it Acknowledgments--}
This work was supported by Grant-in-Aid for Scientific Research from JSPS, KAKENHI Grant Number JP21H05017, and partly by North Campus Instrumental Analysis Station, Kyoto University. K. U. acknowledges support from JSPS KAKENHI Grant Numbers JP22K18322, JP22K03484, JP25K22008, JP26K21749.
S. T. acknowledges support from JSPS KAKENHI Grant Numbers JP26K00662-1 and JP24K06891. Y. M. acknowledges support from JSPS KAKENHI Grant Numbers JP24H00191, JP25K07235, JP26K00646, JP26H01281. 
\end{acknowledgments}

\bibliography{apssamp}

\clearpage
\appendix

\onecolumngrid
\begin{center}
  \textbf{\large Supplemental Material for}\\[2pt]
  \textbf{\large ``Prominent Dimensional Effects on High-order Harmonic Generation in \\ Strongly Correlated Electron Systems''}
\end{center}
\twocolumngrid

\section{Details of the experimental setup}
The experimental setup used in this manuscript is the same as that used in Ref.~\cite{Uchida2022PRL}.
A Ti:sapphire regenerative amplifier (pulse width: 35 fs, pulse energy: 7 mJ, center wavelength: 800 nm, repetition rate: 1 kHz) was used as the laser source. Approximately 4 mJ of the output was used to generate the MIR driving field. Signal beams centered at 1180 nm and 1570 nm were first produced using a dual optical parametric amplifier system (Light Conversion TOPAS-TWINS), and MIR pulses (center wavelength: 4.8 $\mu$m) were then generated by difference-frequency mixing in a GaSe crystal. Residual input signals were blocked with a long-pass filter (cutoff wavelength: 4 $\mu$m). The polarization angle and intensity of the MIR pulses were controlled using three wire-grid polarizers and a liquid-crystal variable retarder (Thorlabs LCC1113-MIR). The MIR beam was reflected by an indium tin oxide (ITO) plate, which has high reflectivity in the MIR and high transmissivity in the visible range. The beam was then focused onto the sample using a reflective objective lens with an effective focal length of 13 mm (working distance: 24 mm). The focal spot size was estimated to be 27 $\rm{\mu m}$ (FWHM) using a knife-edge measurement, and the pulse duration was estimated to be 100 fs via electro-optic sampling. Samples were mounted in a cryostat with a 1-mm-thick $\rm{CaF_2}$ window, and their position was adjusted with mechanical stages to align with the center of the MIR spot. Reflected high-harmonic emission was collected by the same objective and transmitted through the ITO plate. The emission was spectrally resolved with a spectrometer and detected using an InGaAs line detector for the third harmonic or a Si CCD camera for higher harmonics.

\section{Sample preparation}
\begin{figure}[t]
 \centering
   \hspace{-0.cm}
   \vspace{0.0cm}
\includegraphics[width=70mm]{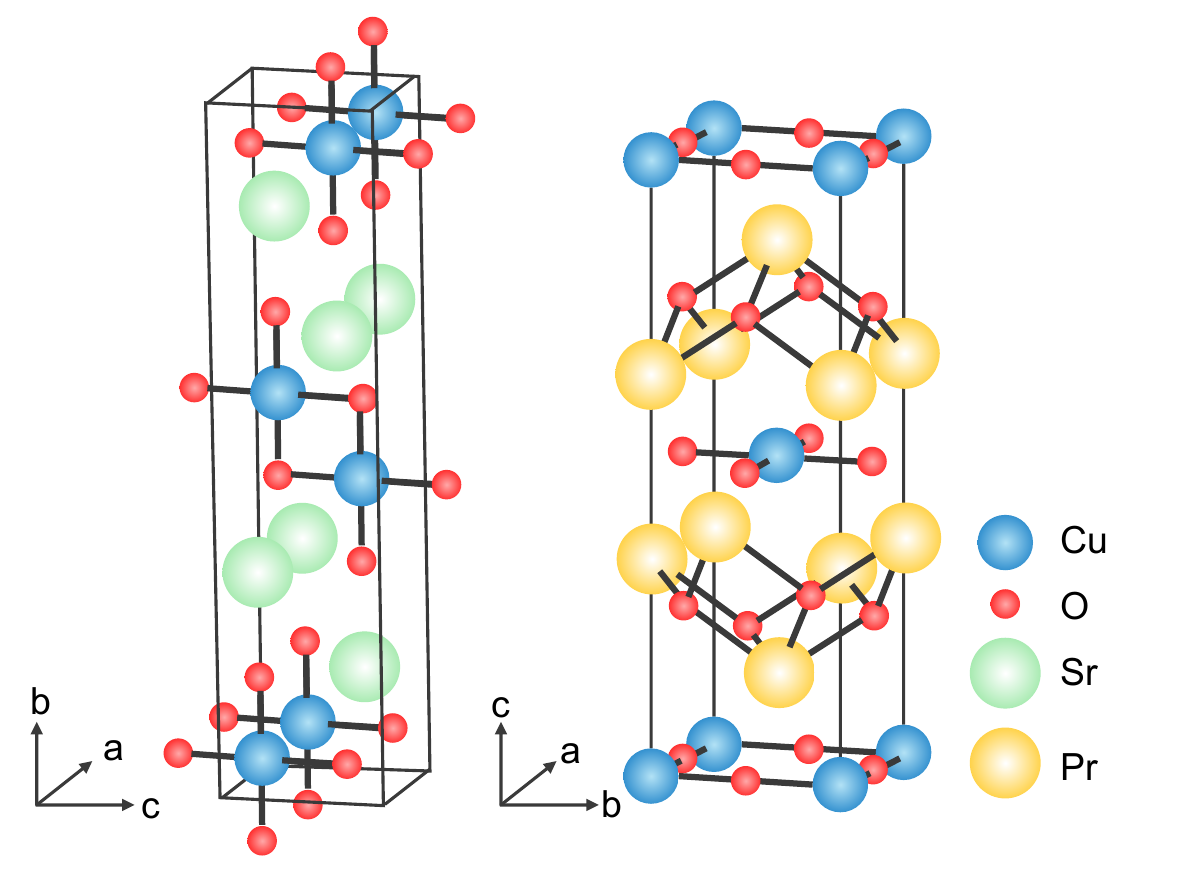} 
 \caption{Crystal structures of $\rm{SrCuO_2}$(left panel) and $\rm{Pr_2CuO_4}$ (right panel). }
 \label{fig:crystal}
\end{figure}
Figure~\ref{fig:crystal} shows the crystal structures of $\rm{SrCuO_2}$ and $\rm{Pr_2CuO_4}$. $\rm{SrCuO_2}$ consists of double Cu-O chains along $c$-axis, on the other hand, $\rm{Pr_2CuO_4}$ has a $\rm{T'}$ structure with $\rm{CuO}$ layers within $ab$-plane. Single crystals of $\rm{SrCuO_2}$ and $\rm{Pr_2CuO_4}$ were grown by means of CuO self-flux method via traveling-solvent floating-zone technique. Poly crystalline feed rods of each compound were obtained by solid reaction with $\rm{SrCO_3}$, $\rm{Pr_6O_{11}}$, and CuO precursors. For the HHG measurement, the grown crystal rods were aligned and cut along the chain and Cu-O bonding direction as well as out-of-chain and out-of-plane direction, respectively, using Laue X-ray diffraction.

\section{Extended experimental data}

\subsection{Driving field intensity dependence}

\begin{figure}[t]
 \centering
   \hspace{-0.cm}
   \vspace{0.0cm}
\includegraphics[width=90mm]{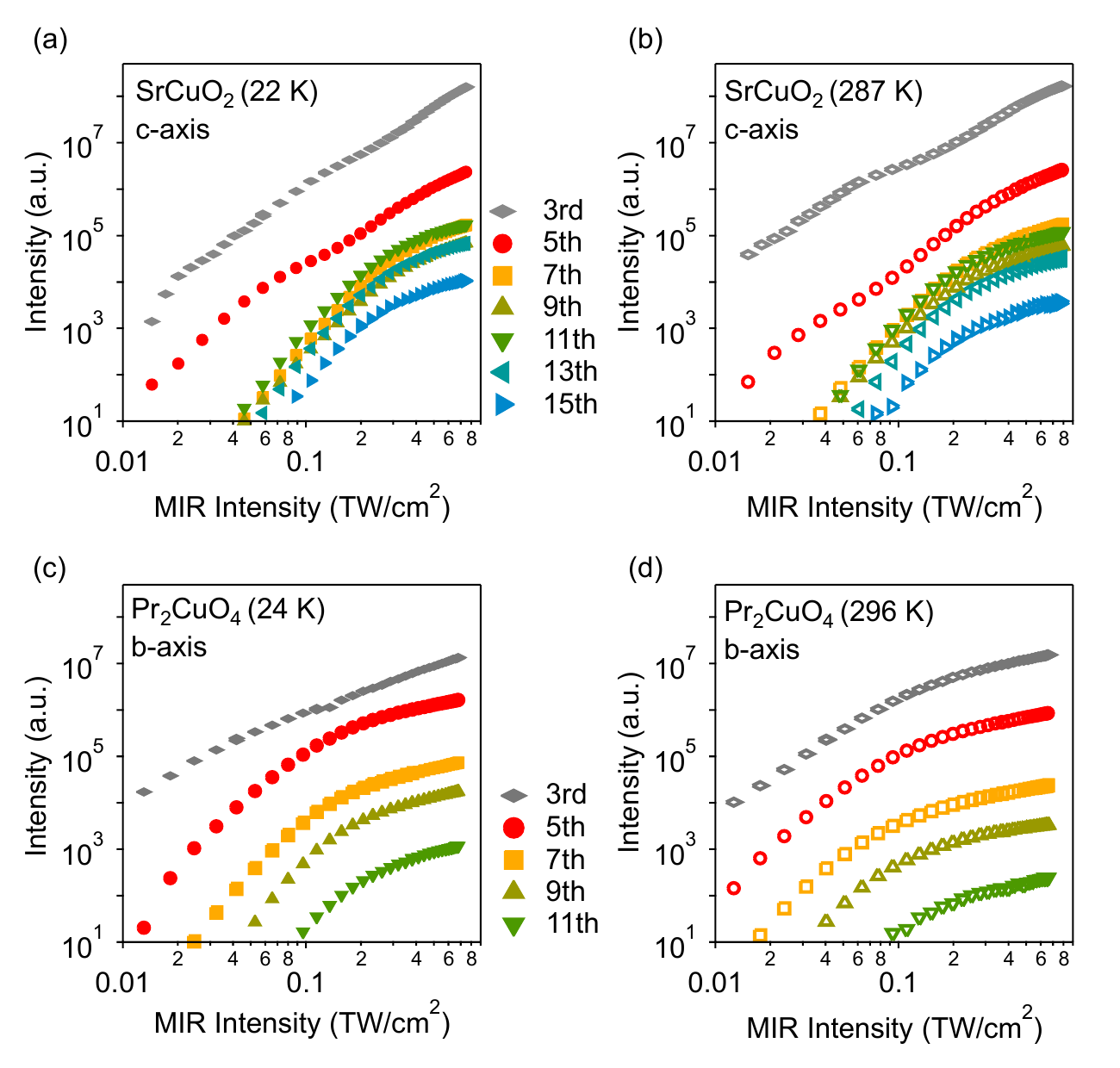} 
 \caption{Driving intensity dependence of high harmonic yields in (a) $\rm{SrCuO_2}$ at 22 K, (b) at 287 K, (c) $\rm{Pr_2CuO_4}$ at 24 K, and (d) at 296 K.}
 \label{fig:Intdep}
\end{figure}

In the main text, we focused on the high-harmonic generation measurements performed under sufficiently strong driving-field conditions, namely at intensities above 0.4~$\rm{TW/cm^2}$. In this section, we present the driving-field intensity dependence of the high-harmonic yields of $\rm{SrCuO_2}$ and $\rm{Pr_2CuO_4}$, measured at low temperature ($\sim 20$~K) and at room temperature (290-300~K) as shown in Figs.~\ref{fig:Intdep} (a)-(d).

In $\rm{SrCuO_2}$, the intensity dependence shows a clear contrast between the lower-order harmonics, namely, the third and fifth harmonics, and the seventh and higher-order harmonics. The emitted photon energies of the third and fifth harmonics are below the gap, whereas those of the seventh and higher-order harmonics are above the gap energy (1.4 eV).

The high harmonic yields of the third and fifth harmonics gradually increase with increasing driving intensity, accompanied by non-monotonic oscillatory structures. By contrast, the seventh and higher-order harmonics tend to increase monotonically at low intensities and then exhibit pronounced saturation at driving intensities above 0.3~$\rm{TW/cm^2}$.

The oscillatory structures in the below-gap harmonics may indicate interference between multiple generation pathways, for example, nonlinear current generated by the propagation of D-H pairs and that arising from D-H pairs polarization.

The universal scaling behavior observed for the above-gap harmonics suggests that these harmonics are governed by a common HHG mechanism, which should be related to D-H-pair recombination. 

The high-harmonic yields in $\mathrm{SrCuO_2}$ exhibit only weak temperature dependence in their driving-intensity scaling, particularly for the above-gap harmonics.

In $\mathrm{Pr_2CuO_4}$ (Figs.~\ref{fig:Intdep}(c) and (d)), except for a slight nonmonotonic behavior of the third harmonic at low temperature, the fifth and higher-order harmonics show similar driving-intensity scaling, with a monotonic increase followed by saturation above 0.2~$\mathrm{TW/cm^2}$.

The high-harmonic yields after saturation tend to become smaller as the temperature increases. Moreover, the pre-saturation intensity scaling is steeper at low temperature, implying that the high-harmonic generation efficiency in the weak-intensity limit is higher at elevated temperature. Although the origin of this behavior remains unclear, it may be related to the enhanced tunneling probability at higher temperature, where the gap energy is reduced. This suggests that, in the weak-intensity regime prior to pronounced saturation, the difference in tunneling probability may play an important role in determining the high-harmonic generation efficiency.

\subsection{Crystal orientation dependence}
\begin{figure}[t]
 \centering
   \hspace{-0.cm}
   \vspace{0.0cm}
\includegraphics[width=95mm]{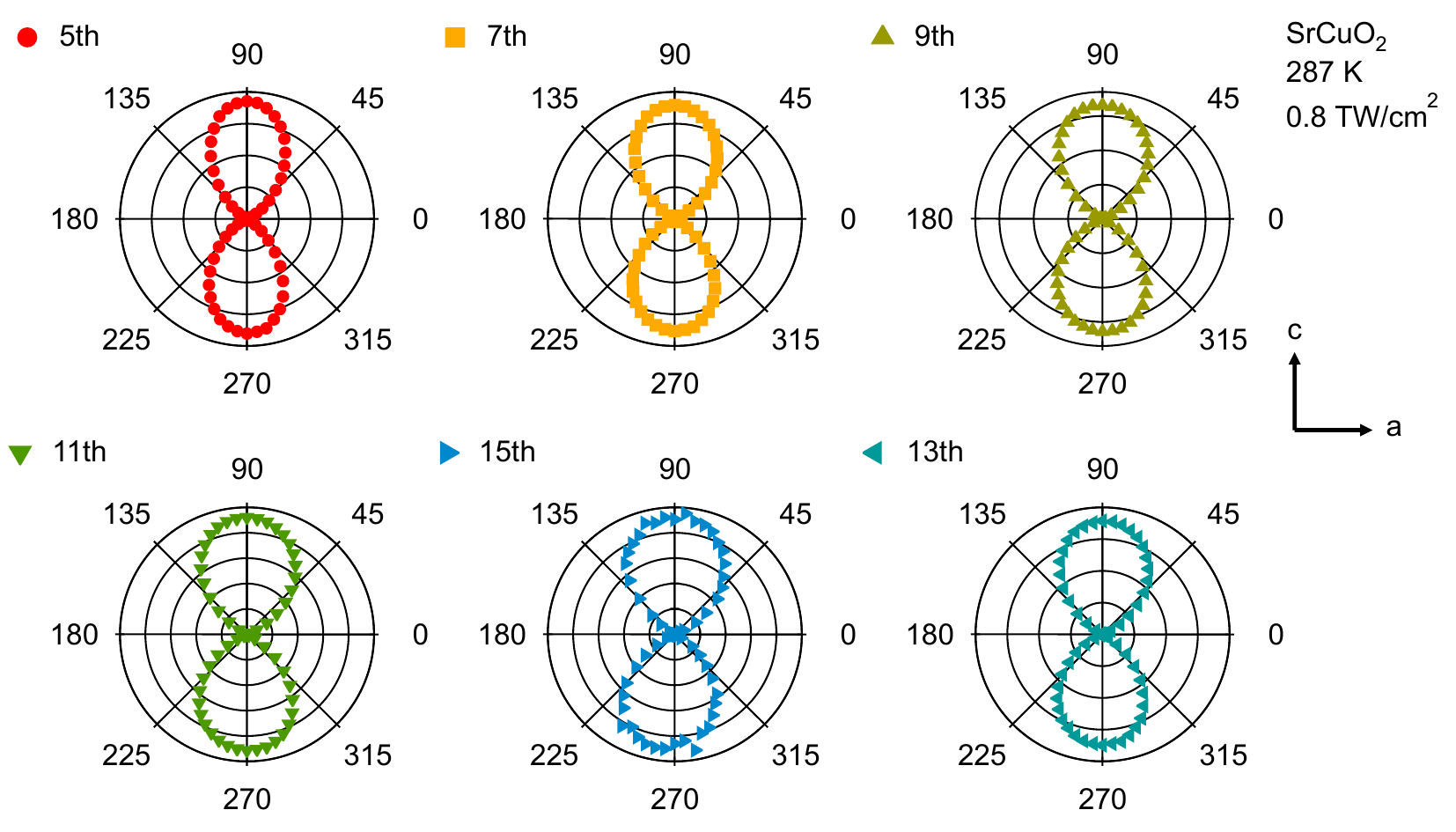} 
 \caption{Crystal orientation dependence of high-harmonic yields in $\rm{SrCuO_2}$ at 287 K. The angle is defined with respect to the $a$-axis of the crystal.}
 \label{fig:OriSCO}
\end{figure}

\begin{figure}[b]
 \centering
   \hspace{-0.cm}
   \vspace{0.0cm}
\includegraphics[width=90mm]{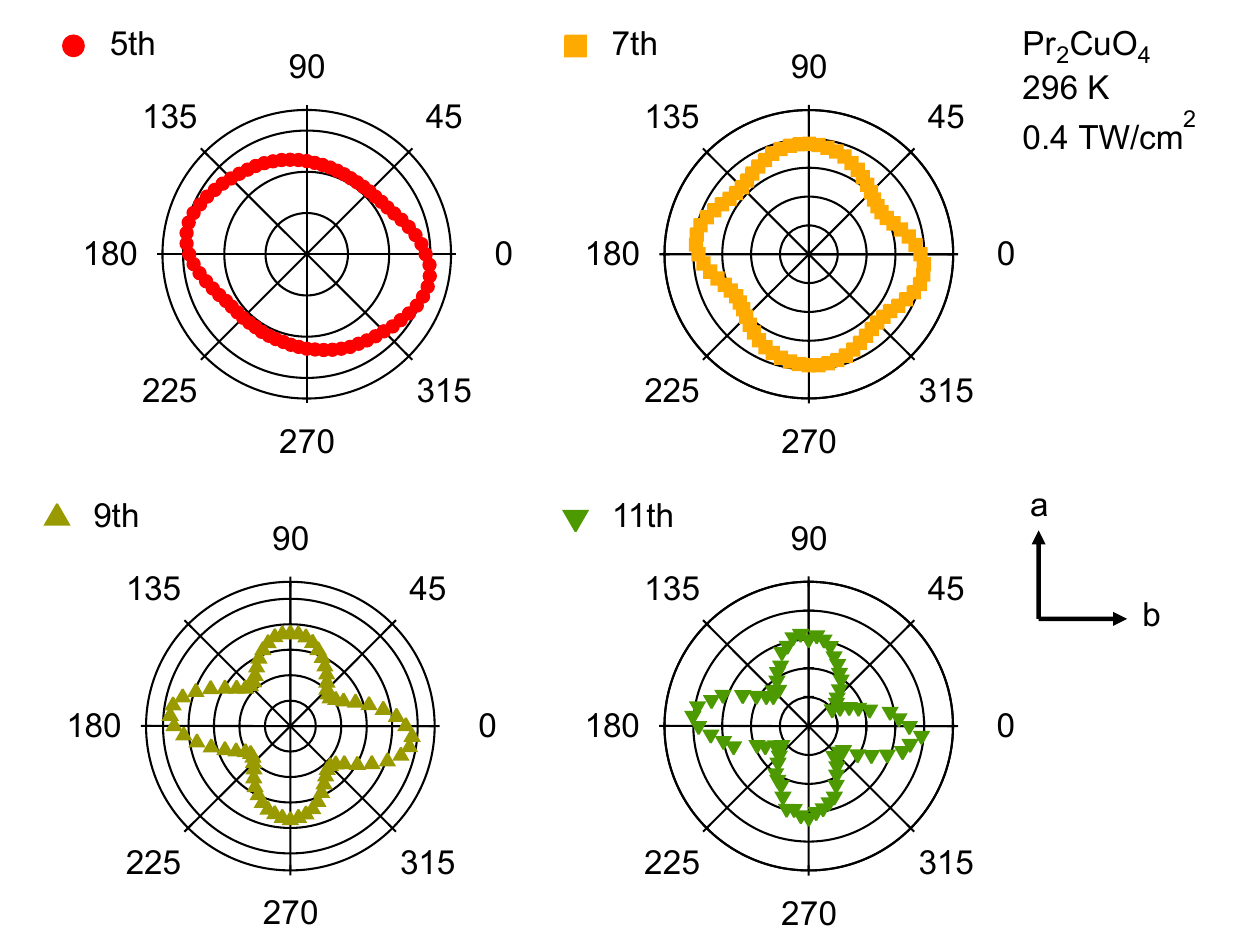} 
 \caption{Crystal orientation dependence of high-harmonic yields in $\rm{Pr_2CuO_4}$ at 296 K. The angle is defined with respective to $b$-axis of the crystal.}
 \label{fig:OriPCO}
\end{figure}
Figure~\ref{fig:OriSCO} shows the crystal-orientation dependence of the high-harmonic yields in $\mathrm{SrCuO_2}$. The high-harmonic yields reach their maxima for driving-field polarization along the $c$-axis, while they are nearly suppressed for polarization along the $a$-axis. The crystal-orientation dependence is nearly independent of harmonic order and can be approximately accounted for by the driving-field component projected onto the crystal $c$-axis. This result reflects the quasi-one-dimensional electronic structure of $\mathrm{SrCuO_2}$ along the $c$-axis.

Figure~\ref{fig:OriPCO} shows the corresponding crystal orientation dependence in $\rm{Pr_2CuO_4}$. Compared with the quasi-one-dimensional $\mathrm{SrCuO_2}$, high-harmonic emission is observed in $\mathrm{Pr_2CuO_4}$ for driving-field polarization along all directions within the two-dimensional CuO plane. This result is consistent with its two-dimensional electronic structure. The high-harmonic yields exhibit stronger anisotropy for higher harmonic orders, with local maxima appearing when the driving field is polarized along the $a$- and $b$-axes. These axes correspond to the nearest-neighbor Cu--Cu directions in the CuO plane.

\section{Details of the Hubbard model}

The physics of cuprates such as $\mathrm{SrCuO_2}$ and $\mathrm{Pr_2CuO_4}$
is well described by the single-band Hubbard model~\cite{Dagotto1994RMP,Tokura_RMP}.
More precisely, undoped cuprates are classified as charge-transfer (CT) insulators.
In these materials, the Cu $3d$ band is split into the lower and upper Hubbard bands,
while the O $2p$ band lies between them, and the Fermi level is located between
the O $2p$ band and the upper Hubbard band.
In principle, this situation can be described by the so-called
$d$–$p$ model, which explicitly includes the Cu $3d$ and O $2p$ orbitals.
However, previous theoretical and experimental studies have demonstrated that
the hole band in CT insulators and the lower Hubbard band in Mott insulators
share essentially the same physical characteristics~\cite{Dagotto1994RMP,Tokura_RMP,Ono2004PRB,Shinjo2021PRB,Okamoto2019SciAdv}.
In particular, the correlated charge dynamics associated with the hole band in CT insulators is closely analogous to that of the lower Hubbard band in Mott insulators, and this fact justifies a description based on an effective single-band Hubbard model through the identification of the O $2p$-derived hole band with the lower Hubbard band.

\begin{figure}[t]
 \centering
   \hspace{-0.cm}
   \vspace{0.0cm}
\includegraphics[width=65mm]{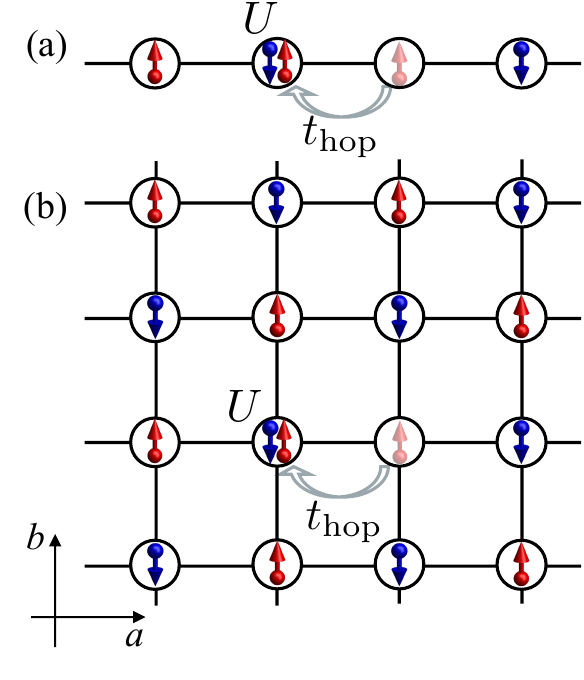} 
 \caption{Schematic illustrations of the (a) 1D and (b) 2D Hubbard models. }
 \label{fig:Hubbard}
\end{figure}

Following this rationale, we also employ the single-band Hubbard model:
\eqq{
\hH(t) = -t_{\rm hop}\sum_{\langle i,j\rangle,\sigma} e^{i {\bf A}(t)\cdot {\bf r}_{ij}} \hc^\dagger_{i\sigma} \hc_{j\sigma}  + U \sum_j \hn_{j\uparrow} \hn_{j\downarrow}, \label{eq:H_origin}
}
where $\hc^\dagger_{i\sigma}$ is the creation operator for an electron with spin $\sigma$ at site $i$, $\langle ij\rangle$ indicates a pair of nearest-neighbor sites, and 
$\hn_{i\sigma} = \hc^\dagger_{i\sigma} \hc_{i\sigma}$. 
$t_{\rm hop}$ is the hopping parameter, 
$U$ is the onsite Coulomb interaction and $e^{i \bA(t)\cdot {\bm r}_{ij}}$ represents the Peierls substitution for the light-matter coupling. 
Here, $\bA(t)$ is the vector potential and ${\bf r}_{ij}$ is the vector from the $j$ site to the $i$ site. 
The charge of the electron is set to unity. 
The electric field is related to the vector potential by $\bE(t) = -\partial_t \bA(t)$.
We focus on half-filled systems in the Mott insulating regime.
We consider the systems on the one-dimensional (1D) chain and the two-dimensional (2D) square lattice (see Fig.~\ref{fig:Hubbard}), which correspond to $\mathrm{SrCuO_2}$ and $\mathrm{Pr_2CuO_4}$, respectively.
For the 1D system, we use the infinite time-evolving block decimation (iTEBD), see Sec.~\ref{sec:iTEBD}.
For the 2D system, we use the nonequilibrium dynamical mean-field theory (DMFT), see Sec.~\ref{sec:DMFT}.
In both case, we set $t=0$ as the initial time.

For the simulations, we use a continuous-wave (CW) pump and a Gaussian-pulse pump:
\eqq{
\bA_{\rm CW}(t) &= \frac{\bE_0}{\Omega} \left[1-\cos(\Omega t)\right],\\
\bA_{\rm Gauss}(t) &= \frac{\bE_0}{\Omega} \, F_{\rm Gauss}(t-t_0,\sigma_0)\, \sin\!\left[\Omega(t-t_0)\right].
}
Here, $\bE_0$ denotes the amplitude of the electric field, $\Omega$ is the pump frequency, and
$F_{\rm Gauss}(t,\sigma_0)=\exp[-t^2/(2\sigma_0^2)]$.
Although the Gaussian pump more closely resembles experimental conditions, we primarily employ the CW pump because finite-temperature iTEBD simulations are limited to relatively short time scales. 
HHG spectra essentially reflect the radiation accumulated during the time-periodic dynamics of the driven system. Thus, when the system rapidly reaches such time-periodic dynamics, short-time simulations under CW excitation can capture the essential features of the HHG process. In the following, we demonstrate that this condition is satisfied for the present parameter set.

The HH intensity is evaluated from the light-induced current ${\bm J}(t)$. 
The current operator $\hat{J}_\alpha$ ($\alpha=a,b$) is defined as
$\hat{J}_\alpha(t)=
-\frac{\partial \hat{H}({\bm A})}{\partial A_\alpha}
\big|_{{\bm A}={\bm A}(t)}.$
The electromagnetic radiation from the system is associated with the acceleration of charge, i.e., the time derivative of the current,
$\dot{\bm J}(t)\equiv \partial_t {\bm J}(t)$.
If the full time evolution under a pump pulse is accessible, the total HH spectrum is expressed as
$I_{\rm HHG}(\omega)=| \dot{\bm J}(\omega) |,$
where ${\bm J}(\omega)$ and $\dot{\bm J}(\omega)$ denote the Fourier transforms of ${\bm J}(t)$ and $\dot{\bm J}(t)$, respectively.
In practice, numerical simulations can be performed in the limited time interval, and the beginning and end of the simulated time trace may contain transient or boundary effects. We therefore extract the HHG spectrum by applying a finite time window to the current, so as to suppress artifacts from the edges of the simulation interval. In particular, for the short-time CW-pump simulations used here, we employ a box window that isolates the radiation emitted within a specified time range:
\eqq{F_{\rm box}(t)=f_{\rm box}\!\left(\frac{t-t_c}{\tau},\eta\right)}
with
\eqq{
f_{\rm box}(x,\eta)
=
\frac{1}{2}\left[1
+
\tanh\!\left(-\frac{2x-1}{2\eta}\right)
\tanh\!\left(\frac{2x+1}{2\eta}\right)\right].
}
This window function smoothly selects the time range
$t \in [t_c-(\tau/2),\,t_c+(\tau/2)]$
with a smoothing width $\eta\tau$.
We then evaluate the HHG spectrum as
$I_{\rm HHG}(\omega)
= | \dot{\bm J}_{\rm Window}(\omega) |^2,$
where $\dot{\bm J}_{\rm Window}(\omega)$ is the Fourier transform of
$F_{\rm box}(t)\dot{\bm J}(t)$.
We define the intensity of the $n$-th order HH peak as $I_{n}\equiv \int_{(n-0.5)\Omega}^{(n+0.5)\Omega} I_{\rm HHG}(\omega) d\omega$.
Furthermore, to analyze the subcycle spectrum, we apply a windowed Fourier transform,
$\dot{\bm J}(\omega,t_p)
=
\int dt\,
e^{i\omega t}\,
F_{\rm Gauss}(t-t_p,\sigma_p)\,
\dot{\bm J}(t),$
which provides the emission profile around time $t_p$ as $I(\omega,t_p)=|\dot{\bm J}(\omega,t_p)|^2$. Here, $\sigma_p$ is the width of the Gaussian window, which is taken much smaller than the period of the pump.

For a systematic comparison among experiments, DMFT simulations, and iTEBD simulations,
we fix the bandwidth of the noninteracting system to $W = 4$ and the Coulomb interaction
to $U = 7$.
This choice is motivated by the fact that $\mathrm{SrCuO_2}$ and $\mathrm{Pr_2CuO_4}$ exhibit similar gap sizes.
We then set the excitation frequency $\Omega = 0.8$ in order to make the ratio
between $\Delta_{\rm Mott}$ and $\Omega$ comparable to that of the experiment.
To evaluate $I_{\rm HHG}$ for the CW pump, we set $t_c=14,\tau=12,\eta=0.15$, which covers radiation from three half-cycles.
For the sub-cycle analyses, we set $\sigma_p=0.7$.

\section{Non-equilibrium dynamical mean-field theory}\label{sec:DMFT}

Dynamical mean-field theory (DMFT) is a theoretical framework based on the Green’s function formalism, and it can efficiently capture correlation effects both in and out of equilibrium~\cite{Georges1996,Aoki2013,Murakami2025RMP}.
The central assumption of DMFT is that the self-energy $\Sigma_{\sigma, ij}(t,t')$ of the Green’s function
\begin{equation}
G_{ij,\sigma}(t,t') \equiv -i\,\langle \mathcal{T}_{\mathcal C} \hat c_{i\sigma}(t)\hat c_{j\sigma}^\dagger(t')\rangle
\end{equation}
is local in space, namely,
\begin{equation}
\Sigma_{ij,\sigma}(t,t') = \delta_{ij}\,\Sigma_{ii,\sigma}(t,t'),
\end{equation}
while fully retaining the local dynamical correlations.
In practice, the self-energy is obtained by mapping the original lattice model onto an effective quantum impurity model coupled to a self-consistently determined bath.
This approximation becomes exact in the infinite-dimensional limit, and DMFT is generally believed to provide a reliable description of strongly correlated systems in high dimensions.
Indeed, its validity for nonequilibrium dynamics has been confirmed by ab initio comparisons between nonequilibrium DMFT and cold-atom quantum simulators for three-dimensional systems~\cite{Kilian2019PRL}.

In this work, we employ nonequilibrium DMFT formulated on the so-called L-shaped Keldysh contour, which consists of one imaginary-time (Matsubara) branch and two real-time branches.
The initial equilibrium state at finite temperature is prepared on the Matsubara branch. We use the nonequilibrium DMFT to simulate the time-evolution of the Hubbard model on the two-dimensional square lattice, allowing the formation of the antiferroagnetic order. Here, the effective impurity model is solved with the non-crossing approximation (NCA)~\cite{Eckstein2010b}, which is known to yield reliable results in the strong-coupling regime. Within DMFT, $\langle S_{z} S_{z}\rangle$ is evaluated as $m_z^2$ with $m_z = |n_{i,\uparrow}-n_{i,\downarrow}|/2$.

\begin{figure*}[tb]
 \centering
   \hspace{-0.cm}
   \vspace{0.0cm}
\includegraphics[width=170mm]{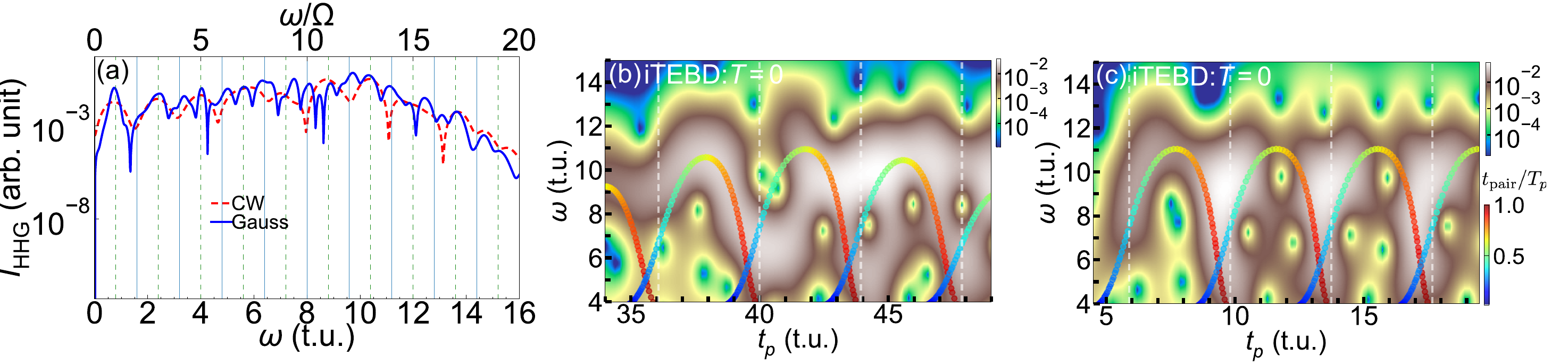} 
 \caption{iTEBD simulations of the HH spectra and the subcycle spectra at $T=0$.
(a) Comparison of the HH spectra obtained under Gaussian-pulse excitation and CW excitation.
(b),(c) Corresponding subcycle spectra for the CW excitation (b) and the Gaussian-pulse excitation (c).
The multicolored dots indicate the emission energy at time $t_p$ within the doublon-holon three-step model.
The vertical dashed lines indicate the times with $A(t)=0$.
We set $W=4$, $U=7$, $\Omega=0.8$, and $E_0=1.2$. For the Gaussian pulse, $t_0=40$, and $\sigma_0=10$. For the subcycle analysis, we use $\sigma_p=0.7$.}
 \label{fig:iTEBD_T0}
\end{figure*}

\section{Finite temperature infinite time evolving block decimation}\label{sec:iTEBD}
We investigate the non-equilibrium dynamics of the 1D Hubbard model in the thermodynamic limit at finite temperature by employing the infinite Time-Evolving Block Decimation (iTEBD) algorithm, which is based on the matrix product state (MPS) formalism~\cite{Vidal2003PRL,Vidal2007PRL}. To handle the thermal mixed state, we utilize the purification method, which maps the mixed state described by the density operator $\hat{\rho}$ acting on the physical Hilbert space $\mathcal{H}_{\mathrm{P}}$ to a pure state $\ket{\Psi}$ defined on an enlarged Hilbert space $\mathcal{H}_{\mathrm{P}} \otimes \mathcal{H}_{\mathrm{A}}$~\cite{Verstraete2004PRL}. Here, $\mathcal{H}_{\mathrm{A}}$ represents an auxiliary, also known as ancilla, Hilbert space that is a copy of the physical space. The pure state in the enlarged Hilbert space $\ket{\Psi} \in \mathcal{H}_{\mathrm{P}} \otimes \mathcal{H}_{\mathrm{A}}$ is constructed such that the density operator is recovered by tracing out the auxiliary degrees of freedom, satisfying the relation $\hat{\rho} = \mathrm{Tr}_{\mathrm{A}} (\ket{\Psi} \bra{\Psi})$.

The simulation is performed according to two procedures: an imaginary time evolution to prepare the thermal state, followed by a real time evolution to simulate the dynamics. We begin by initializing the system at infinite temperature ($\beta = 0$), where the density operator is proportional to the identity operator. In the purification picture, this corresponds to a direct product of maximally entangled pairs between each physical state and its corresponding auxiliary state. In the Hubbard model studied in this paper, the Hilbert space on each site is represented in the basis of four states $\ket{0},\ket{\uparrow},\ket{\downarrow},\ket{\uparrow\downarrow}$, where $0$, $\uparrow$, and $\downarrow$ represents an empty site, an electron with the up spin, and an electron with the down spin, respectively. Then the initial pure state is constructed as 
\begin{align}
  \ket{\Psi_{0}}=\prod_{j}\frac{1}{2}
    \big(&\ket{0}_{\mathrm{P},j}\ket{0}_{\mathrm{A},j}
      +\ket{\uparrow}_{\mathrm{P},j}\ket{\uparrow}_{\mathrm{A},j}
\nonumber\\
      &+\ket{\downarrow}_{\mathrm{P},j}\ket{\downarrow}_{\mathrm{A},j}
      +\ket{\uparrow\downarrow}_{\mathrm{P},j}\ket{\uparrow\downarrow}_{\mathrm{A},j}\big).
\end{align}
This is a purification of the mixed state at the infinite temperature, 
\begin{align}
  \hat{\rho}_{0}=\prod_{j}\frac{1}{4}
    \big(&\ket{0}_{j}\bra{0}_{j}
      +\ket{\uparrow}_{j}\bra{\uparrow}_{j}
\nonumber\\
      &+\ket{\downarrow}_{j}\bra{\downarrow}_{j}
      +\ket{\uparrow\downarrow}_{j}\bra{\uparrow\downarrow}_{j}\big).
\end{align}

The thermal mixed state at a finite temperature $\beta = 1/(k_{\mathrm{B}} T)$ is represented as $\hat{\rho}_{\beta}=e^{-\beta \hat{H}(0)}\hat{\rho}_0$, and it corresponds to the purified state $\ket{\Psi_{\beta}} = (e^{-\beta \hat{H}(0)/2} \otimes \hat{1}_{\mathrm{A}}) \ket{\Psi_{0}}$. In fact, we can confirm 
$\hat{\rho}_{\beta}=\mathrm{Tr}_{\mathrm{A}}\ket{\Psi_{\beta}} \bra{\Psi_{\beta}}$. The imaginary time evolution of the pure state in the extended Hilbert space can be calculated through the usual procedure of iTEBD. 

Once the system reaches the desired temperature $\beta$, we start the real-time evolution to study the dynamic properties of the system. We apply the unitary time evolution operator 
$\hat{U}(t) = \mathcal{T}  e^{-i\int_{0}^{t}\hat{H}(t')dt'}$ to the thermalized state $\ket{\Psi_{\beta}}$, where $\mathcal{T}$ represents the time-ordering operator, thus we obtain
\begin{align}
  \ket{\Psi_{\beta}(t)}=(\hat{U}(t) \otimes \hat{1}_{\mathrm{A}}) \ket{\Psi_{\beta}}.
\end{align}
The real time evolution is implemented by iTEBD similarly to the imaginary time evolution. At the time $t$, the time-dependent expectation value of a physical observable $\hat{O}$ is computed as
\begin{align}
  \braket{\hat{O}(t)} =& 
    \braket{\Psi_{\beta}(t) | (\hat{O} \otimes \hat{1}_{\mathrm{A}}) | \Psi_{\beta}(t)} \\
    =& \braket{\Psi_{\beta} | (\hat{U}^{\dagger}(t) \hat{O} \hat{U}(t) \otimes \hat{1}_{\mathrm{A}}) | \Psi_{\beta}} \nonumber\\
    =& \mathrm{Tr} (\hat{U}^{\dagger}(t) \hat{O} \hat{U}(t) \hat{\rho}_{\beta}). \nonumber
\end{align}

In the iTEBD calculation, we treat the Hubbard model as the spinless fermion ladder by regarding the spin-up and spin-down as the chain index of the ladder. The action of the real and imaginary time evolution operators is decomposed by the fourth order Trotter formula, and we employ the swap gate method to avoid the action of the long-range operators~\cite{Stoudenmire2010NJP}. We utilize two conserved quantum numbers: the total number of spin-up fermions and that of spin-down fermions, to reduce the computational cost.

\section{Extended numerical results}
\begin{figure}[t]
 \centering
   \hspace{-0.cm}
   \vspace{0.0cm}
\includegraphics[width=50mm]{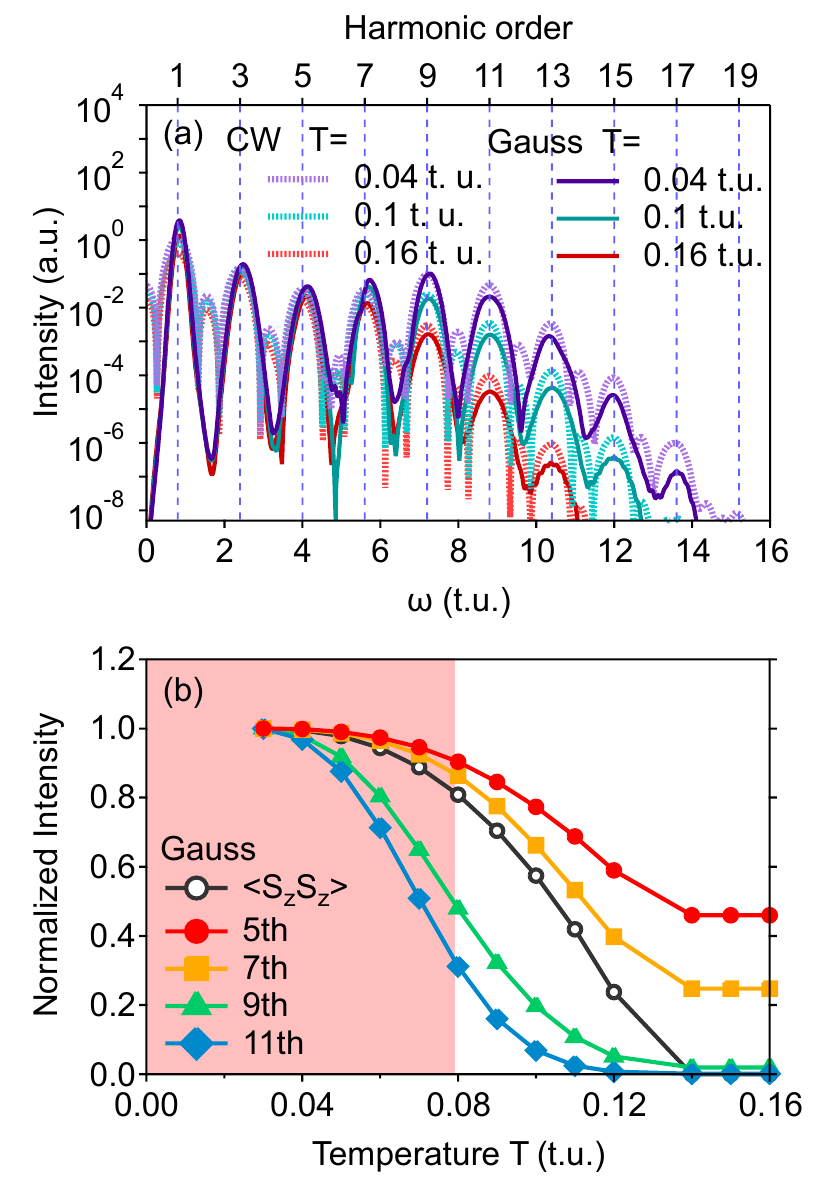} 
 \caption{DMFT simulations of the HH spectra under Gaussian-pulse excitation.
(a) Temperature dependence of the HH spectra for the Gaussian-pulse excitation (solid lines) and the CW excitation (dashed lines).
For the CW case, a box window with $t_c=14$, $\tau=12$, $\eta=0.15$ is used.
(b) Temperature dependence of the $n$th peak intensity ($I_n$) and the nearest-neighbor spin--spin correlation $\langle S_z S_z \rangle$.
 We set $W=4$, $U=7$, $\Omega=0.8$, $E_0=1.2$, $t_0=40$, and $\sigma_0=10$. The electric field is along the $a$-axis.}
 \label{fig:IHHG_DMFT_comprare}
\end{figure}

\begin{figure}[t]
 \centering
   \hspace{-0.cm}
   \vspace{0.0cm}
\includegraphics[width=50mm]{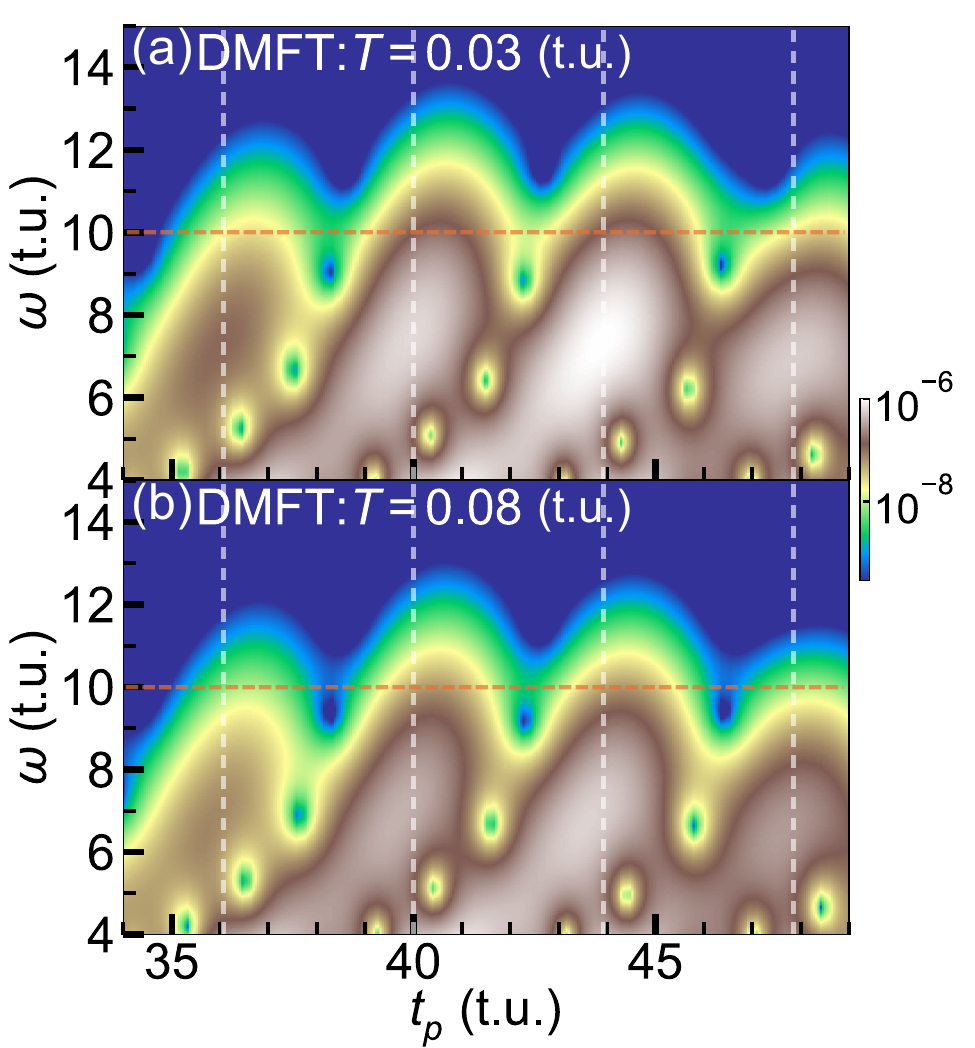} 
 \caption{DMFT simulations of the subcycle spectra under Gaussian-pulse excitation. (a) is for $T=0.03$ and (b) is for $T=0.08$.
 We set $W=4$, $U=7$, $\Omega=0.8$, $E_0=1.2$, $t_0=40$, and $\sigma_0=10$. The electric field is along the $a$-axis.
 For the subcycle analysis, we use $\sigma_p=0.7$. The vertical dashed lines indicate the times with $A(t)=0$.}
 \label{fig:Subcycle_DMFT_comprare}
\end{figure}
\subsection{Comparison between long-time and short-time simulation}

In the main text, to systematically explore the temperature dependence of the HH spectra, 
we apply a CW pump and use a Fourier window to focus on the radiation over a few cycles of the pump. 
This procedure is adopted because, at finite temperature, the iTEBD method allows us to simulate only relatively short-time dynamics. 
To justify this approach, we compare the HH spectra evaluated with the CW pump, as in the main text, 
to those obtained from longer-time simulations with a Gaussian pump, 
using the conventional iTEBD at $T=0$ and the nonequilibrium DMFT.

Figure \ref{fig:iTEBD_T0} shows the results of the iTEBD simulation at $T=0$. 
The generic features of the HH spectra, such as the range of intense radiation, 
are nearly the same for the CW and Gaussian pumps (see Fig.~\ref{fig:iTEBD_T0}(a)). 
Similar features can also be found in the subcycle spectra (see Figs.~\ref{fig:iTEBD_T0}(b),(c)). 
Namely, the evolution of the radiation spectra roughly follows the prediction of the doublon–holon (D-H) three-step model~\cite{Murakami2021PRB}.
In particular, strong radiation occurs due to the recombination of D-H pairs with a long time interval between their creation and recombination, 
corresponding to the red-colored markers in Figs.~\ref{fig:iTEBD_T0}(b),(c). 
For both the CW and Gaussian pumps, some signals deviate from the prediction of the D-H three-step model. 
This deviation can be attributed to the relatively high excitation frequency compared to the hopping parameter, 
as well as the relatively strong field intensity used in the simulations.

We now turn to the DMFT simulations. 
In Fig.~\ref{fig:IHHG_DMFT_comprare}(a), we directly compare the HH spectra obtained from the CW and Gaussian pump simulations. 
The generic features of the HH spectra are very similar over a range of temperatures.
In Fig.~\ref{fig:IHHG_DMFT_comprare}(b), we show the temperature dependence of the $n$th peak intensity ($I_n$) and the nearest-neighbor spin--spin correlation $\langle S_z S_z \rangle$ for the Gaussian pump. 
This can be directly compared with Fig.~3(d) in the main text for the CW pump. 
We again find that the overall temperature dependence of the peak intensity are qualitatively the same.
In Fig.~\ref{fig:Subcycle_DMFT_comprare}, we show the subcycle spectra for the Gaussian pump, which corresponds to Figs.~4(c)(d) in the main text for the CW pump.
We again find the good qualitative agreement.

\subsection{Polarization dependence of high-harmonic intensity in two-dimensional systems}
\begin{figure}[b]
 \centering
   \hspace{-0.cm}
   \vspace{0.0cm}
\includegraphics[width=70mm]{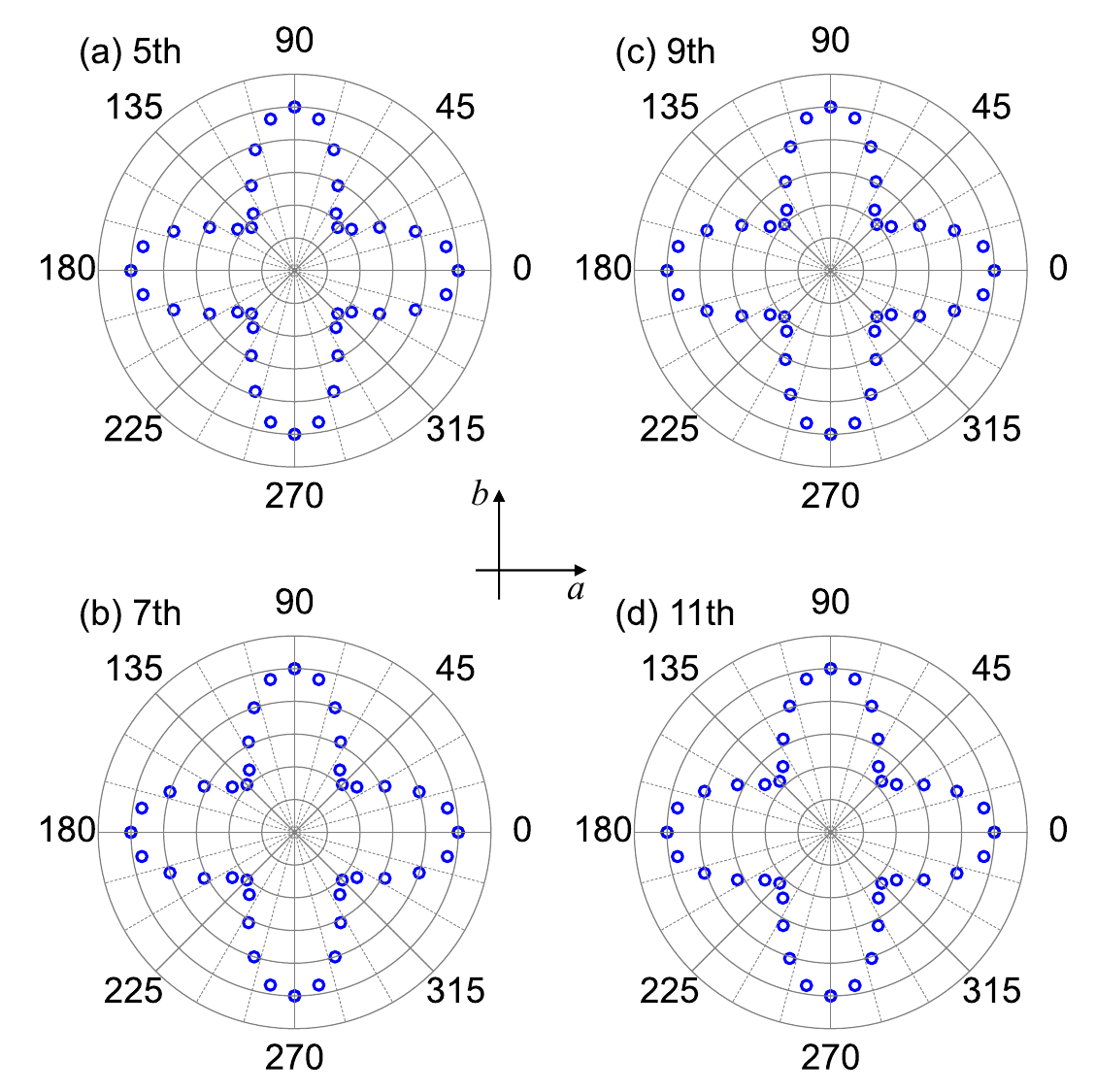} 
 \caption{Polarization dependence of the $n$th peak intensity ($I_n$) for the DMFT simulations under Gaussian-pulse excitation.
We set $W=4$, $U=7$, $T=0.04$, $\Omega=0.8$, $E_0=1.2$, $t_0=40$, and $\sigma_0=10$.}
 \label{fig:PolDep_DMFT}
\end{figure}

The simulation of the 2D Hubbard model using nonequilibrium DMFT successfully reproduces the polarization dependence of high-harmonic generation in $\mathrm{Pr_2CuO_4}$, see Fig.~\ref{fig:PolDep_DMFT}.
The high-harmonic intensity is maximized when the driving field is polarized along the bond direction, i.e., along the $a$- or $b$-axis.

\end{document}